# Can AI automatically analyze public opinion?

# A LLM agents-based agentic pipeline for timely public opinion analysis

Jing Liu[1]*, Xinxing Ren[2]*,Yanmeng Xu[1], Zekun Guo[3]†

[1]Brunel Design School, Brunel University of London

[2]College of Engineering, Design and Physical Sciences, Brunel University of London

[3]Faculty of Science and Engineering, University of Hull

*These authors contributed equally to this work.

†Corresponding author: Z.Guo2@hull.ac.uk

## Abstract

This study proposes and implements the first LLM agents–based agentic pipeline for multi-task public opinion analysis. Unlike traditional methods, it offers an end-to-end, fully automated analytical workflow without requiring domain-specific training data, manual annotation, or local deployment. The pipeline integrates advanced LLM capabilities into a low-cost, user-friendly framework suitable for resource-constrained environments. It enables timely, integrated public opinion analysis through a single natural language query, making it accessible to non-expert users. To validate its effectiveness, the pipeline was applied to a real-world case study of the 2025 U.S.–China tariff dispute, where it analyzed 1,572 Weibo posts and generated a structured, multi-part analytical report. The results demonstrate some relationships between public opinion and governmental decision-making. These contributions represent a novel advancement in applying generative AI to public governance, bridging the gap between technical sophistication and practical usability in public opinion monitoring.

## Introduction

In recent years, public opinion analysis has developed rapidly, driven by advances in data analysis technologies[1]. Scholars generally apply these analytical technologies to address individual core tasks in public opinion analysis, such as data collection, sentiment analysis, and topic extraction[2, 3]. Therefore, although public opinion research traditionally falls within the field of mass communication, its analytical studies are situated within the domain of data analysis.

Specifically, advances in technical methods for public opinion analysis have progressively enhanced the precision and scope of individual analytical tasks. For example, classical methods employed machine learning (ML), such as support vector machine (SVM) and decision tree, primarily for basic sentiment classification[4, 5]. The emergence of deep learning (DL), including convolutional neural network (CNN)[6] and recurrent neural network (RNN)[7], has enabled models to capture more complex linguistic patterns and contextual nuances, thereby improving performance in both sentiment analysis[8] and opinion extraction[9]. However, despite these advancements, the application of such technical methods still faces several limitations. Traditional ML models rely heavily on hand-crafted features and struggle to capture the complex semantics of unstructured and informal language [10, 11]. These models typically require labor-intensive annotation and iterative model tuning, leading to low efficiency[12]. Although DL models have improved accuracy, they often lack interpretability, functioning as “black boxes” whose outputs are difficult to justify [13].

Moreover, DL methods depend on large-scale, high-quality labeled datasets [14], which are often costly and challenging to obtain.

Therefore, although various technical methods, from ML to DL, have been employed to address core tasks in public opinion analysis, the overall analytical process still faces several challenges. These include the inherent complexity of the technologies, a strong reliance on labor-intensive annotation and high-quality labeled datasets for domain adaptation, and the technical expertise required to operate them. More critically, the above techniques typically concentrate on individual tasks, such as sentiment analysis or topic extraction, without integrating them into a unified analytical workflow[15]. This fragmentation of analytical processes, along with the absence of a comprehensive view of public opinion, impedes the development of end-to-end solutions and further reinforces dependence on task-specific techniques. As a result of these challenges, the core tasks involved in public opinion analysis process often require specialized programming knowledge, substantial computing resources, and manual labor, creating significant barriers for non-expert users[16, 17].

In practical governance scenarios, decision-making based on public opinion analysis often lacks timeliness[18]. Some scholars point out that one reason is officials and elites tend to selectively respond to portions of public opinion based on their prior subjective experiences, particularly those that align with their own interests, rather than addressing the broader needs of society[19]. Others suggest that governments, particularly the Chinese government, prefer to use retrospective public opinion data after related social events have concluded, as a cautious and risk-averse approach[20]. Another practical challenge in public opinion analysis is the lack of adequate human and material resources. The establishment of computing infrastructure required for timely, large-scale public opinion analysis is generally expensive and unaffordable for many local governments. Furthermore, professional analysts are needed to operate these computing infrastructures effectively.

To address both technical and practical challenges, this research envisions a possibility: if a tool could be developed that is low-cost, user-friendly, timely, and has low technical barriers for integrated public opinion analysis, could it help resolve current challenges such as dataset dependence, limited human and material resources, the need for technical expertise, lack of timeliness, and technological complexity?

Fortunately, recent advances in artificial intelligence, particularly in large language models (LLMs), make such a tool feasible. LLMs based on transformer architectures (e.g., BERT, GPT-series) [21] has significantly advanced the depth and flexibility of public opinion analysis, particularly in multidimensional opinion detection[22], often surpassing earlier methods in analytical accuracy[23]. In addition, LLMs demonstrate enhanced capabilities in contextual understanding and deep semantic interpretation, enabling more accurate analysis of nuanced and complex public opinion content[24]. LLMs support for prompt-based task execution allows users to interact through natural language instructions, reducing the technical barrier for non-expert users[25]. Moreover, LLMs' ability to perform zero-shot and few-shot learning significantly reduces the reliance on large labeled datasets[26]. These characteristics make LLM-powered systems not only more powerful but also more accessible and user-friendly. Building on these advantages, the use of LLM agents has become an effective way to apply LLMs. LLM agents are built upon LLMs and therefore inherit strengths such as contextual understanding, prompt-based task execution, and low data dependence [27, 28]. By assigning each agent a specific task, multiple agents can work in combination to execute diverse tasks simultaneously or sequentially [29, 30]. This modular structure enables efficient, multi-task processing within short time frames [31]. Furthermore, since agents operate through simple API calls to pre-trained LLMs[32], they greatly reduce the need for human and material resources as well as for expert-level programming, making them highly cost-effective and accessible to a wider range of users.

Based on the advantages of LLM agents, this research designs an LLM agents–based agentic pipeline, deployed on remote servers, for integrated multi-task public opinion analysis. The core tasks incorporated into this pipeline are identified based on the typical analytical flow of public opinion analysis[2], which consists of data collection, sentiment analysis, topic extraction, and report generation. This AI analytical pipeline tool is intended to address current challenges in the public opinion analysis process. By encapsulating core tasks of public opinion analysis into modular components, using a Coordinator Agent's built-in utility for data collection and prompt-initialized LLM agents for sentiment analysis, topic extraction, and report generation, this pipeline delivers a fully automated, end-to-end workflow. Users only need to input a single natural language query, after which each agent executes its assigned task automatically with no further user involvement. This design significantly reduces the implementation complexity of public opinion analysis technologies and lowers barriers to adoption for users. In addition, this form of multi-task automated execution constitutes an integrated public opinion analysis solution, unlike earlier technical methods that executed each analytical task in isolation. The tool runs via OpenAI API invocations, so users need to submit requests and receive results, no local GPU setup or compute management is required, thereby greatly reducing dependence on hardware resources and manual effort. To the best of our knowledge, this is the first research to implement an LLM-based agentic pipeline for public opinion analysis.

The key contributions of this research are as follows: 1) It presents the first LLM agents-based agentic pipeline specially designed for multi-tasks public opinion analysis. 2) It is the first tool that integrates data collection, sentiment analysis, topic extraction, and report generation into an integrated, end-to-end process. 3) It does not rely on domain-specific, manually labeled datasets and labor-intensive annotation. 4) It lowers technical barriers by offering a fully automated, user-friendly experience. 5) It operates under a low-cost architecture suitable for deployment in resource-limited environments. 6) It supports timely public opinion analysis, enabling governments to adopt this tool as an alternative to selective responses based on subjective experience and delayed use of retrospective public opinion data. Collectively, these contributions represent a novel attempt to bridge advanced AI analytical capabilities with the needs of public opinion analysis.

## Designing the LLM agents-based agentic pipeline

Building on the description of the proposed pipeline tool discussed above, this section presents the architecture and operational process of this LLM agents-based tool for public opinion analysis. As Figure 1 shows, once the user submits a query, each agent automatically executes its assigned task in sequence, passing its output file as the input to the next stage, thereby forming a fully automated, end-to-end workflow.

It should be noted that all LLM inference is performed via OpenAI API invocations by each agent on remote servers, so no local GPU or compute setup is required (see Appendix A for implementation environment details).

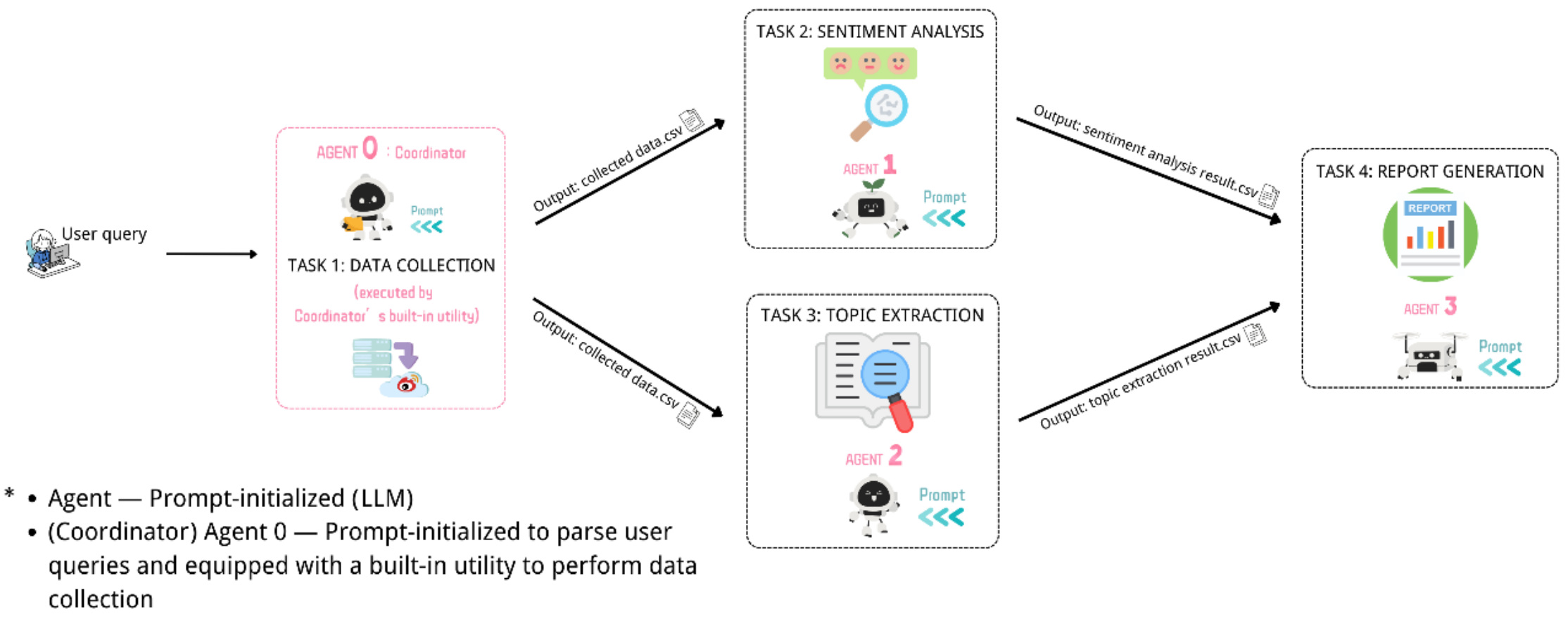

**Figure 1 The LLM agents-based agentic pipeline**

## 1. Agent 0 (Coordinator): parse user query & data collection

Agent 0 serves as the entry point for the entire pipeline. Upon receiving a single natural language query from the user, containing three key pieces of information: “keyword”, “time span”, and “social media platform”, Agent 0 is initialized with its prompt and invokes the LLM to extract and normalize this input into a structured JSON object with “event_keywords”, “start_date”, “end_date”, and “social media platform”.

Next, Agent 0 passes this JSON to its built-in data collection module, which launches a specially developed crawler to automatically collect matching public opinion posts and export them to a CSV file (see Appendix B for implementation details). This CSV is then passed to downstream LLM agents for sentiment analysis and topic extraction, completing the first step of the fully automated workflow.

It should be noted that the pipeline currently supports only Sina Weibo (one of the most popular social media platforms in China, like Twitter), as the built-in crawler is specifically tailored to Weibo’s interface. The crawler will be described in detail in the next section.

## 2. Agent 1: sentiment analysis task

After the data collection stage produces a CSV file containing the collected posts, the pipeline invokes Agent 1, which is initialized with a dedicated sentiment analysis prompt.

First, the pipeline reads the CSV and extracts each post’s text field. Agent 1 then combines its prompt, which defines its role as a sentiment classifier and specifies the desired output format, with the extracted post texts. The LLM processes each entry and returns a structured response, consisting of sentiment labels such as positive, neutral, or negative. Agent 1 parses this response, aligns each label with the corresponding original post, and writes the results to a new CSV file as the output of the sentiment analysis stage.

This resulting file, containing one sentiment label per post, is then passed directly to the downstream report generation agent, ensuring that the entire sentiment analysis step is fully automated and requires no further human intervention.

## 3. Agent 2: topic extraction task

Agent 2 is also invoked by the pipeline to analyze the collected data CSV file, similar to Agent 1. It is initialized with a dedicated topic extraction prompt that defines its role and specifies the expected output structure.

Following the same execution process as Agent 1, the pipeline reads the CSV file, extracts the content of each post, and combines these texts with Agent 2's system prompt. The LLM analyzes each entry to identify one or more salient topics (e.g., vaccine safety, public trust, side effects) and returns them in a structured format. Agent 2 then parses the responses, maps each list of topics back to its corresponding post, and writes the consolidated results to a new CSV file, which serves as the output of the topic extraction stage.

This output, detailing the detected topics for each post, is subsequently passed to the report generation agent, thereby completing the topic analysis stage of the end-to-end workflow.

## 4. Agent 3: report generation task

After the sentiment analysis and topic extraction stages generate two corresponding CSV files, the pipeline invokes Agent 3, which is initialized with a report generation prompt that instructs the agent to synthesize the merged sentiment and topic outputs into a cohesive summary.

The pipeline reads both CSV files, merges their rows into a single data payload, containing each post's text, its sentiment label, and its extracted topics, and passes this consolidated input, along with Agent 3's prompt, to LLM. Agent 3 then generates the final public opinion report based on this information.

Finally, Agent 3 parses LLM's output and exports the completed report, thereby concluding the end-to-end public opinion analysis workflow.

## 5. Formalization of this agentic pipeline

The end-to-end workflow of the proposed public opinion analysis pipeline can be concisely represented as:

$$U \rightarrow A_{0(parse+collect)} \rightarrow \{\, A_1 \parallel A_2 \} \rightarrow A_3$$

Here, U denotes the user's single natural language input query. $A_{0(\text{parse+collect})}$ represents the Coordinator Agent, which parses U into JSON parameters via its prompt and then uses its built-in data collection utility to generate a CSV file containing the collected data. $\{A_1||A_2\}$ refers to the Sentiment Analysis Agent $A_1$ and the Topic Extraction Agent $A_2$, which operate in parallel on the collected data CSV to produce two separate output files: one for sentiment analysis and one for topic extraction. $A_3$ is the Report Generation Agent, which takes both CSV files as input and synthesizes them into a final cohesive public opinion report.

This formal notation directly maps each stage of the pipeline to its corresponding agentic operation, illustrating the modular and coordinated structure of the system.

## 6. Specific prompts for each agent

In the pipeline, five dedicated prompt files, each stored as a plain-text TXT, define the behavior of every agentic component in the public opinion workflow. Figure 2 illustrates the shared prompt execution cycle: at initialization, an agent loads its TXT prompt, uses it to parse its input (user query for Agent 0 or upstream CSV files for Agents 1–3), and serializes the model's response into JSON or CSV for the next stage,thus enforcing a clean, modular, fully automated workflow.

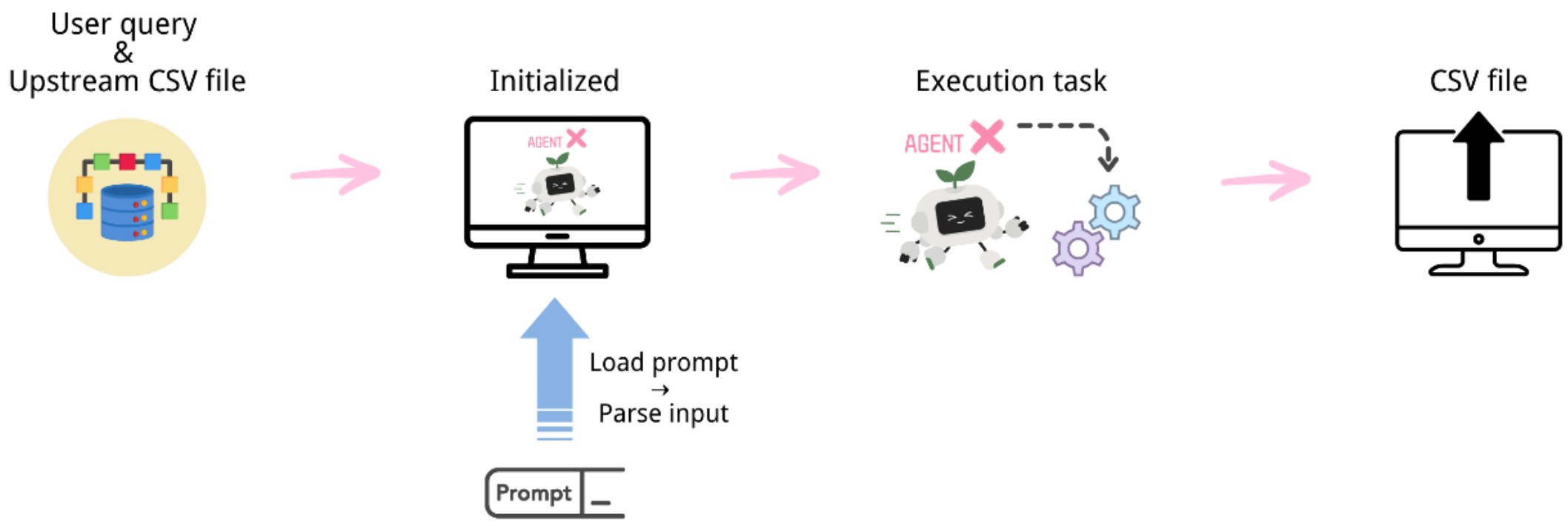

* • User query is fed into Agent 0 (Coordinator), which parses it into JSON instructions.
• The resulting CSV is then fed into downstream agents (A1–A2) for analysis, whose outputs in turn feed Agent 3.

**Figure 2 Prompt-driven agent initialization and execution workflow**

Full listings and explanations of each agent's prompt are provided in Appendix C.

## Case study: timely public opinion analysis on the 2025 U.S. tariff topic

To validate the applicability of the proposed LLM agents-based agentic pipeline, this research applied it to one of the most prominent and currently trending public opinion topics: the U.S.–China tariff dispute. This topic was triggered by the U.S. government's announcement of new tariffs on Chinese goods on April 9, 2025, and was immediately followed by a countermeasure from the Chinese government. On Sina Weibo, the topic attracted high public discussion, engaging participants across a wide spectrum, including official media, celebrities, self-media accounts, and ordinary users, who joined discussion immediately as the topic emerged.

This tariff topic was selected for the case study based on the following key considerations.

First, the topic emerged during the period in which this research was being conducted, providing an opportunity to verify the pipeline's timeliness. Therefore, the proposed agentic pipeline was applied to analyze public opinion data within the initial 24-hour window of the tariff topic, as it was executed on April 9, 2025—the very day the topic first emerged.

Second, the public discussion related to the U.S. tariff topic on that day generated a large volume of content, as evidenced by the topic's high popularity trend. According to a topic monitoring tool on Sina Weibo, Trending Topic Engine[33], the tariff topic reached a popularity score of 2.77 million on April 9, 2025. As shown in Figure 3, this popularity reflects intense user engagement during the initial 24-hour period. In addition, because the tariff topic concerns national interests, the related discussions inevitably involved strong nationalist sentiment. These analytical aspects, namely sentiment intensity and topic popularity, align closely with the capabilities of the proposed agentic pipeline, which includes sentiment analysis and topic extraction. This suggests that the tariff topic is particularly well-suited for integrated public opinion analysis, enabling the execution of a fully automated, end-to-end analytical workflow that produces a relatively comprehensive report.

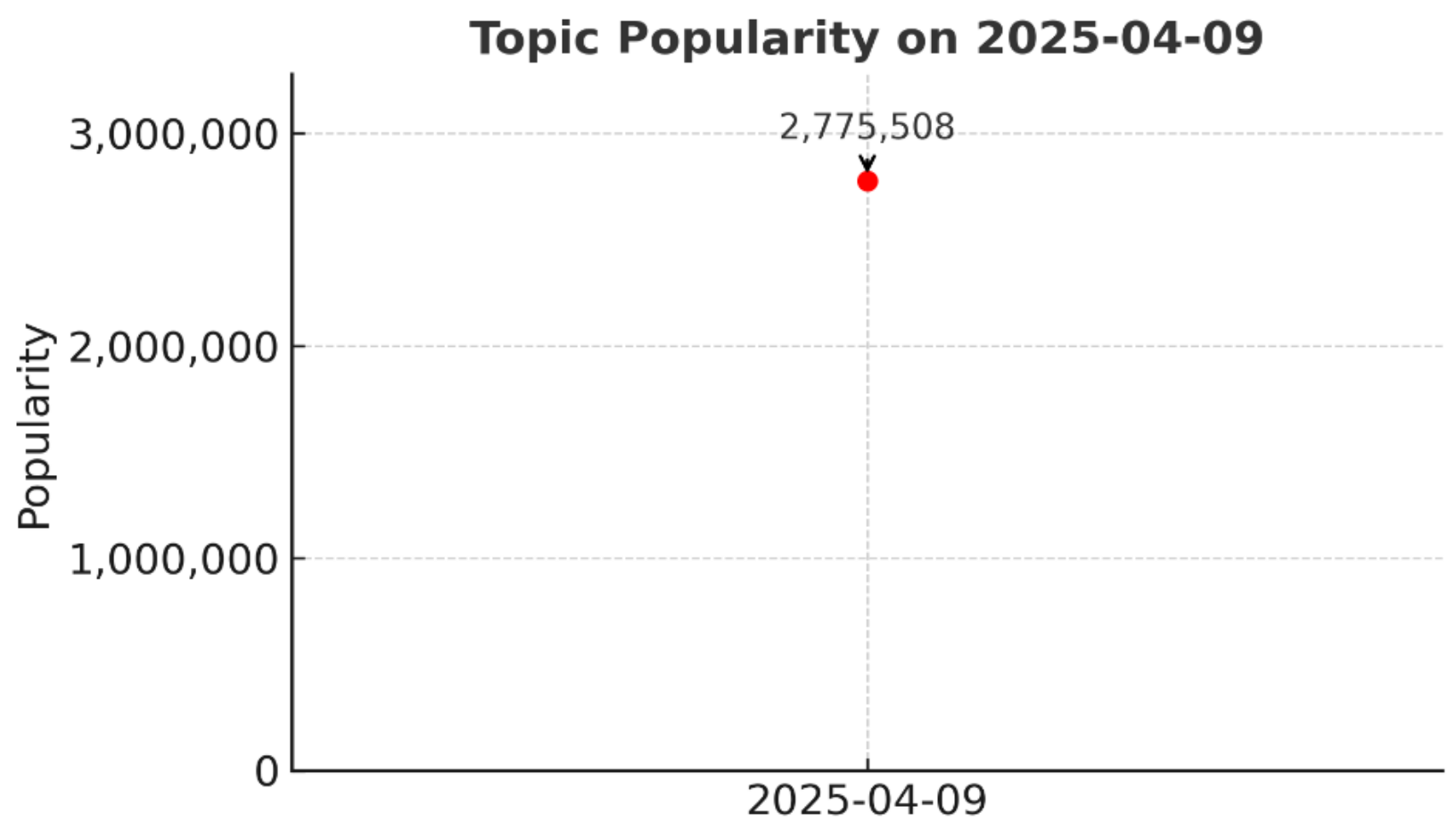

**Figure 3 The traffic topic's popularity**

Third, while the public discussion was still evolving, a series of concrete policy responses to this topic were already issued by both the Chinese and U.S. governments. This suggests the potential value of using public opinion analysis to examine the degree of correspondence between public concerns and the content of decision-making. For instance, if the government's responses substantially align with the major concerns reflected in public opinion, it may indicate that public opinion can play a significant role in influencing the decision-making process. Moreover, if the interaction between public opinion and decision-making can be verified in the context of this tariff topic, it would further support the notion that early-stage public opinion analysis can contribute to more proactive policy related responses. This is particularly relevant in this case, as concrete governmental responses were observed during the early phase of the tariff topic's public opinion.

## 1. Parse user query and timely data collection

The analysis begins when the user submits a single natural language instruction:

“Please analyze public opinion on Sina Weibo about the U.S. tariff on April 9, 2025” or the shorthand “U.S. tariff, 2025.4.9, Weibo.” Upon receiving this instruction, Agent 0 (Coordinator) is then initialized with its dedicated prompt and immediately invokes the LLM to extract and normalize three key parameters—event_keywords: “U.S. tariff”; start_date and end_date: “2025.4.9”; and event_release_platform: “Sina Weibo”—into a structured JSON configuration.

Without further manual intervention, Agent 0 then passes these parameters to its built-in data collection utility. This module launches the specially developed Sina Weibo crawler in this research, which wraps the keyword in hashtags (in this case study, transforming “U.S. tariff” into “#U.S. tariff#”), splits the 24-hour window from 00:00 4/9/2025 to 00:00 4/10/2025 into hourly segments, issues paginated HTTP requests up to the Weibo platform’s 50-page limit, and extracts each post text, timestamp, user ID, source device, and engagement metrics (Repost, Comment, and Like counts). Any posts outside the specified window or lacking the hash tagged keyword are discarded.

In total, 1,572 Weibo posts were collected as the initial wave of public opinion. This dataset is written to a CSV file named “collected data” and then automatically handed off to downstream LLM agents for sentiment analysis and topic extraction, with no further user input required. A representative sample of this CSV is shown in Table 1.

**Table 1 Sample of the collected data**

| User ID | Timestamp | Source | Reposts/Comments/ | Post Text |
| --- | --- | --- | --- | --- |

| | | Device | Likes | |
|---|---|---|---|---|
| Science Future Person | April 9, 00:10 | Frank's Android | 6/1/6 | #Who is hurt by the U.S. tariff policy# A senior official from Trump's… |
| Bengbu Daily | April 9, 08:57 | Weibo Video Account | 0/1/1 | #Tariff Policy Sparks Widespread Concern Across the U.S.# The U.S. has announced… |

## 2. Automated sentiment analysis

Once the CSV file of collected data is available, the Sentiment Analysis Agent (Agent 1) is automatically invoked to execute sentimental classification on the collected dataset. To optimize both the accuracy and efficiency of this sentiment analysis task, batch size tuning experiments are required. These experiments showed that grouping posts into fixed batches of ten per API invocation yielded the more reliable matches between the model's predicted sentiment labels and the posts' actual sentiment. Consequently, the pipeline enforces this ten-post batch constraint in the Jupyter Notebook's looping logic, striking an effective balance between throughput and classification reliability.

Under the guidance of its dedicated sentiment analysis prompt, Agent 1 classifies each post as positive, neutral, or negative and appends these labels to the per-post records. After labeling each batch, it computes the overall sentiment distribution for the dataset. The Notebook's summary output reported 207 positive, 371 neutral, and 995 negative posts, which in aggregate indicated that public opinion on the U.S.–China tariff topic was predominantly negative. Table 2 illustrates sample output from Agent 1's sentiment classification, showing how each post is labeled with its corresponding sentiment.

Inspecting the exported CSV file of sentiment analysis result revealed a single "mixed" label that fell outside the three predefined categories. Rather than discarding this unexpected value, the proposed agentic pipeline retains it in the final CSV to preserve analytical completeness and surface the LLM's nuanced interpretation of ambiguous content. This discrepancy (1,573 labels counted vs. 1,572 posts) reflects the model's adaptive flexibility in real-world text.

**Table 2 Sample output from sentiment analysis (Agent 1)**

| Batch Number | Post Text | Sentiment |
|---|---|---|
| 2 | #Who is hurt by the U.S. tariff policy# A senior official from Trump's… | negative |
| 4 | #Tariff Policy Sparks Widespread Concern Across the U.S.# The U.S. has announced… | negative |

## 3. Automated topic extraction

Upon completion of sentiment analysis, the Topic Extraction Agent (Agent 2) is invoked automatically to perform topic extraction on the previously collected dataset.

Agent 2 reads the same CSV file of collected data produced by Agent 0 and, to ensure stable performance within LLM token limits, processes the data in fixed-size batches of up to ten posts, mirroring the batching strategy used by the Sentiment Analysis Agent (Agent 1). Guided by its topic-specific prompt, the agent groups each batch into semantically coherent clusters and assigns one or more topic labels to each post based on linguistic similarity and issue relevance. Posts lacking clear topic convergence remain untagged, while those exhibiting strong semantic overlap receive multiple labels.

The raw topic extraction result output is saved as a CSV file, containing 361 unique topic assignments that span broad themes and nuanced subtopics, thus providing a detailed map of the topic content landscape. Table 3 presents representative examples from the extracted results.

**Table 3 Sample output from topic extraction (Agent 2)**

| Batch Number | Post Text | Topics |
|---|---|---|
| 2 | #Who is hurt by the U.S. tariff policy# A senior official from Trump's… | Escalation of the China–US trade war, Responses of major global economies to US tariff increases |
| 3 | #Tariff Policy Sparks Widespread Concern Across the U.S.# The U.S. has announced… | Impact of US tariff policy on corporate product pricing structure, US tariff policy may widen the wealth gap |

A subsequent aggregation step then consolidates these 361 assignments into 22

distinct, high-level topics, capturing central public concerns, such as consumer goods pricing impacts, multilateral trade system challenges, and strategic economic countermeasures. Table 4 summarizes the final 22 high-level topics. This two-tiered topic extraction approach, detailed extraction followed by aggregated summarization, ensures both depth and clarity in the final topic analysis.

**Table 4 Aggregated results: 22 high-level topics**

| Number | Topics | Post Quantity |
|---|---|---|
| 1 | Impact of US tariff on imported consumer goods for ordinary residents | 305 |
| 2 | Escalation of the China–US trade war and China's countermeasures | 338 |
| 3 | Challenges to the global multilateral trading system | 110 |
| 4 | Ripple effects of US tariff policy on the global economy | 409 |
| 5 | Policy suggestions for structural adjustment of China's economy under US tariff | 98 |
| 6 | Support for small and medium-sized enterprises and regulation of e-commerce development | 6 |
| 7 | Education policy adjustments from the perspective of cultural security | 2 |
| 8 | Impact of US tariff on the channels and prices of | 1 |
| | imported pet food | |
| 9 | Response of China's Ministry of Foreign Affairs to US tariff increases | 64 |
| 10 | Disputes over international multilateral trade order under the China–US trade war | 101 |
| 11 | Impact of US tariff policy on corporate product pricing structure | 433 |
| 12 | Responses of major global economies to US tariff increases | 288 |
| 13 | Economic cooperation positions of China-Japan-Korea and China-Europe in tariff disputes | 48 |
| 14 | Energy and tariff negotiations in US–EU trade disputes | 39 |
| 15 | Use of energy exports by the US as a tool of trade pressure | 20 |
| 16 | Countermeasures of the EU against US trade pressure | 60 |
| 17 | US tariff policy may widen the wealth gap | 127 |
| 18 | US tariff increases on multiple countries | 78 |
| 19 | Impact of US tariff on reshoring of manufacturing | 47 |
| 20 | Strategic and security considerations behind US tariff policy | 80 |
| 21 | Impact of US tariff policy on imported consumer goods | 167 |
| | for ordinary residents | |
| 22 | Impact of US tariff policy on reshoring of manufacturing | 48 |

## 4. Timely integrated analytical summary

Once sentiment classification and topic extraction are complete, the Report Generation Agent (Agent 3) is automatically invoked to synthesize a unified natural language report. The agentic pipeline reads both the sentiment analysis result and topic extraction result CSV files, merges their rows into a single data payload, and sends this payload together with the report generation prompt to the LLM.

Under the guidance of this dedicated prompt, the model first calculates the exact counts of posts in each sentiment category (positive, neutral, negative) and tallies the frequency of each extracted topic. It then performs interpretive synthesis by identifying dominant emotional trends, highlighting the most frequently occurring themes, and drawing connections between sentiment and topic patterns.

The final output of Agent 3 is structured into four labeled sections: “Sentiment Overview,” “Topic Distribution,” “Sentiment–Topic Insights,” and “Conclusions & Recommendations”. Each section presents a clear narrative analysis based directly on the computed statistics and clustered topic content. This multi-section report is printed in the Jupyter Notebook and saved as the final output of the pipeline.

# Case study results

## 1. Agent 3: automated multi-task analysis and policy-oriented reporting

Building on the results of sentiment analysis and topic extraction, Agent 3 synthesized a four-part analytical report. The contents below were generated based on its prompt instructions, but also include emergent reasoning beyond the original prompt scope:

### 1.1 Sentiment overview: predominantly negative sentiment driven by economic anxieties

Agent 3 first summarized the sentiment distribution calculated in the earlier stage: among 1 572 posts, 995 (63.3 %) were labeled negative, 371 (23.6 %) neutral, and 207 (13.2 %) positive, with one post (0.1 %) unexpectedly tagged as "mixed".

The report emphasized that negative sentiment, which was primarily associated with rising consumer costs, industrial pressure, and distrust in U.S. policy motives, was the dominant tone in public opinion.

Neutral posts were mostly informational reposts or analytical commentary, while the limited number of positive posts reflected support for China's countermeasures and confidence in domestic policy responses.

### 1.2 Topic overview: economic pressure and trade tensions dominate

Next, Agent 3 presented a high-level summary of 22 aggregated topics, which were derived from an initial set of 361 topics identified in the raw CSV file. The top three themes were global economic ripple effects (409 mentions, 13%), the escalation of the China–U.S. trade war and countermeasures (338, 10%), and the impact on consumer-goods pricing (305, 10%), all of which highlighted a dual focus on macro- and micro-level economic pressures.

Further analysis revealed that discussions on corporate pricing strategies (433 mentions, 14%) even surpassed those related to individual consumer concerns, indicating a heightened public sensitivity toward upstream industrial responses. In addition, topics such as challenges to the multilateral trading system (110, 3%) and strategic security considerations (80, 3%) reflected broader anxieties about the long-term policy environment and national security risks.

These interwoven themes demonstrate not only widespread concern over immediate economic consequences but also deeper public reflection on policy choices and global power dynamics.

### 1.3 Sentiment–topic interaction: everyday concerns evoke stronger sentiments

Agent 3's cross-analysis of sentiment and topic labels revealed a clear pattern. Posts related to consumer-goods pricing and corporate pricing strategies showed the highest proportion of negative sentiment, with more than 70 percent of posts falling into this category. In contrast, discussions involving trade frameworks, multilateral policy planning, and geopolitical strategy were more frequently classified as neutral, with approximately 30 to 40 percent of those posts receiving neutral labels and only a relatively small share identified as negative.

Positive sentiment was most often found in posts that expressed support for China's policy responses and confidence in the country's economic resilience. These made up approximately 15 percent of posts in those topic areas.

This contrast suggests that direct, everyday economic issues, such as rising prices or increased business costs, are more likely to trigger strong sentiment reactions compared to abstract or long-term policy topics. Within the context of the U.S.–China tariff debate, public anxiety appears to stem mainly from concerns related to personal livelihood, while technical and strategic discussions tend to generate more neutral or detached responses. Understanding this difference can help decision-makers communicate more effectively, especially by addressing sentiment salient issues such as consumer protection and cost-of-living support.

## 1.4 Conclusions & Recommendations: enhancing communication, subsidy, and diplomatic responses

Moving beyond descriptive synthesis, Agent 3 autonomously incorporated historical context by comparing the 2025 public discourse with the earlier U.S.–China tariff disputes in 2018. Based on this analysis, it formulated three strategic policy recommendations.

First, it advocated for proactive public communication. This included encouraging authorities to clearly explain tariff mechanisms, implementation timelines, and corresponding mitigation measures. Such efforts are intended to reduce misinformation and alleviate public anxiety.

Second, it recommended the introduction of targeted subsidies or price-stabilization programs for essential consumer goods. These measures directly address cost-of-living pressures, which were identified as key drivers of negative sentiment.

Third, it called for the strengthening of multilateral diplomatic engagement. By promoting constructive dialogue and collaborative initiatives, China and its international partners can enhance the resilience of the global trading system and help prevent further escalation.

This form of emergent policy reasoning, which was not included in the original prompts, illustrates the pipeline's ability to go beyond automated analysis. It demonstrates the potential of LLM-based agentic workflows to produce actionable governance insights and support anticipatory public policy development.

## 2. Revealing the relationship between public opinion and governmental decision-making

The analytical results of the tariff case study, generated through the proposed agentic pipeline, reveal key aspects of the relationship between public opinion and governmental decision-making.

First, there may be a reciprocal relationship between public opinion and governmental decision-making. On April 9, 2025, a substantial surge in negative sentiment was detected by the pipeline, with 63.3% of posts classified as negative and widespread discussion focusing on the economic consequences of the U.S. tariff announcement. On the same day, the Chinese government responded with concrete measures: it imposed an additional 50% duty on U.S. goods, raising the total tariff rate to 84%, and restricted the operations of six U.S. firms in China. The close temporal and thematic alignment between these policy actions and public sentiment suggests a dynamic interplay, where public opinion may have influenced governmental decision-making while also being shaped by them in return.

Second, public opinion analysis can serve as a useful approach for evaluating whether decision-making responds to public concerns. By April 9, the proposed agentic pipeline had identified that over 70% of posts related to consumer goods pricing and corporate pricing strategies conveyed negative sentiment, making these the most sentiment-intense topics in the dataset. On the same day, both the Chinese and U.S. governments issued retaliatory tariff measures that directly addressed these economic concerns. This alignment between public sentiment and decision-making response suggests that sentiment–topic analysis offers a meaningful basis for assessing the responsiveness of decision-making to public priorities.

Third, early-stage public opinion analysis can provide valuable references for timely decision-making. In this case, the agentic pipeline produced a integrated analytical report within the first 24 hours of the tariff topic's emergence. Notably, the report included several policy-oriented recommendations, generated autonomously by Agent 3 without explicit prompting. These suggestions included improving public communication to clarify the impact of tariffs, introducing targeted subsidies to mitigate consumer cost pressures, and strengthening multilateral diplomatic engagement to reduce strategic tensions. The relative alignment between these AI-generated recommendations and the official responses issued later on April 9 suggests that early automated analysis not only captures public sentiment and concerns but can also offer forward-looking insights to inform timely decision-making.

## 3. Additional value: emergent insights and interpretive reasoning in the LLM-agentic pipeline

A particular notable strength of the LLM-agentic pipeline lies in its ability to generate emergent insights that go beyond the explicit instructions embedded in the prompt. While the agents were initialized with relatively general task descriptions, such as performing sentiment classification or extracting topics, the resulting outputs often included deeper interpretive reasoning, contextual associations, and policy-relevant inferences. For instance, Agent 3 not only reported on sentiment and topic patterns, but also synthesized a set of strategic policy recommendations and drew historical comparisons with the 2018 U.S.–China tariff disputes. These outcomes were not directly specified in the prompt, yet emerged from the LLM's capacity to integrate linguistic, contextual, and inferential knowledge during execution.

This emergent behavior illustrates the added value of leveraging LLMs in structured analytical pipelines, as it enables the analysis to uncover policy-relevant patterns and generate governance-oriented insights with minimal human guidance.

## 4. Advantages of the agentic pipeline: delivering timely, automated, and accessible end-to-end analysis

It is precisely the adoption of a prompt-driven, LLM-agents architecture that enables the realization of several critical advantages in public opinion analysis, as demonstrated in the U.S.–China tariff case study. By encapsulating each core analytical task — data collection, sentiment classification, topic extraction, and report generation — into modular agents coordinated through a unified pipeline, a fully integrated, multi-task, end-to-end workflow can be delivered.

First, the agentic pipeline enables timely execution by initiating data collection on April 9, 2025, the exact day the tariff topic emerged, thus capturing its earliest stage. Second, it achieves full automation, with all stages operating independently after a single user query. Third, the system maintains a low technical barrier by executing all LLM functions via remote API invocations, eliminating the need for local model deployment or custom configuration. Finally, the pipeline ensures user-friendliness, as non-technical users are only required to submit one natural language prompt, without writing code or managing infrastructure.

Taken together, these attributes illustrate how the integration of LLM agents transforms public opinion analysis from a fragmented, expertise intensive process into a streamlined and accessible solution. This approach not only improves analytical efficiency and accessibility but also demonstrates the practical viability of deploying LLM agents for scalable, continuous opinion monitoring in real-world governance contexts.

## 5. Minimal human and material resource requirements

Unlike traditional ML and DL methods, which rely on extensive domain-specific pretraining using large, manually annotated datasets, the agentic analytical pipeline developed in this research leverages zero-shot inference through prompt-initialized LLM agents. Each agent directly invokes the OpenAI API, leveraging the linguistic knowledge embedded in the LLM, and performs public opinion analysis without requiring any additional fine-tuning or labeled datasets. This prompt-only approach shifts human effort from labor-intensive data annotation to lightweight prompt engineering, thereby significantly reducing manpower requirements.

Furthermore, since all inference is executed remotely via API, the system eliminates the need for on-premise GPUs or dedicated computational infrastructure. In the U.S.–China tariff case study, a complete end-to-end execution involved only a few hundred API invocations, with a total cost of less than ten US dollars—several orders of magnitude lower than the GPU hours and annotation labor typically required for training and deploying comparable ML or DL models.

Together, these features allow this agentic pipeline to operate under minimal human and material resource constraints, making it highly efficient and broadly deployable.

## Conclusion and discussion

This study designs and implements a novel LLM agents-based agentic pipeline for public opinion analysis. To the best of our knowledge, this is the first research to realize a fully automated, end-to-end analysis workflow—encompassing data collection, sentiment analysis, topic extraction, and report generation—triggered by a single natural language query. This pipeline was validated through a real-world case study of the 2025 U.S.–China tariff dispute, demonstrating its practical feasibility and analytical effectiveness in a time-sensitive context.

Compared with traditional ML and DL methods, the proposed agentic pipeline offers a series of technical and operational advantages. First, it eliminates the need for domain-specific labeled datasets or GPU infrastructure by leveraging prompt-based, zero-shot LLM inference[34]. Second, it significantly lowers the technical barrier for users. Third, the system is cost-efficient and cloud-deployable. Fourth, the multi-agent architecture ensures stable orchestration and seamless handoff between agents.

The empirical findings from the tariff case study further reinforce the pipeline's utility. The analysis revealed that negative sentiment concentrated around pricing issues and corporate policy, and this aligned temporally and thematically with decision-making responses from both the Chinese and U.S. governments. These results support three core insights: 1)Public opinion and decision-making actions may influence each other dynamically. 2)Public opinion analysis can serve as an approach to evaluate whether decision-making reflects public priorities. 3)Early-stage opinion analysis can provide reference for timely and targeted decision-making.

Notably, the agentic pipeline exhibited “emergent reasoning” capabilities: Agent 3 generated forward-looking policy recommendations and historical comparisons not explicitly defined in the prompts[35]. This indicates that LLM agents can move beyond task execution to provide higher-order governance insights, thereby expanding the value of AI in political communication and public management.

However, this research also has limitations. The case study was based on a single-day dataset, limiting longitudinal analysis of public opinion evolution. Additionally, the pipeline currently supports only one platform (Sina Weibo). Future work should extend the observation window, support cross-platform data integration, and develop more adaptive prompt interaction mechanisms to handle complex, evolving public opinion[36]. Future work could also draw on feedback-driven multi-agent LLM frameworks, which model the adaptive interactions between regulators and other actors [37], to simulate how public opinion and governmental decision-making shape each other over time.

Overall, the angentic pipeline developed in this research demonstrates the feasibility of using AI to support timely, automated, and integrated public opinion analysis. It provides a novel analytical tool for enhancing policy responsiveness in the age of AI and opens new avenues for timely, data-informed governance.

## Appendix A. Implementation environment

### 1. Language & Notebook

- Python 3.13.1
- Jupyter Notebook 7.2.2

### 2. Third-party libraries

- openai – LLM API client
- requests – HTTP crawling
- beautifulsoup4 – HTML parsing
- pandas – CSV I/O & data manipulation
- tqdm – Progress bars
- matplotlib – Optional plotting in utils.py

### 3. Standard library modules

- base64
- os
- json
- csv
- re
- datetime
- logging

## Appendix B. Implementation details of the specifically developed crawler for Sina Weibo data collection

The pipeline's data collection step relies on a specially developed crawler tailored to Sina Weibo's search interface. To align with Weibo's hashtag-centric organization of discussion threads, the crawler automatically wraps each user defined keyword in hash symbols (e.g., #Keyword#) before issuing the query, ensuring that topic focused posts are collected.

Under the hood, the crawler divides the user specified time span into one-hour intervals. For each interval, it issues HTTP GET requests,using randomized headers to evade anti-scraping measures,up to Weibo's pagination limit of 50 pages per query. Within each page, it parses the HTML to extract post text, timestamp, user ID, source device, and engagement metrics (likes, reposts, and comments). Posts with timestamps outside the current hourly window or whose text does not contain the exact hashtagged keyword are discarded. On HTTP or parsing errors, the crawler logs the exception, sleeps for a random short delay, and retries the request once before continuing.

Written in Python 3.10 with requests, pandas, and fake_useragent, the crawler accumulates the filtered records and writes them to a structured CSV (Collected_data.csv). This file then serves as the single source of truth for all downstream LLM agents, enabling the rest of the workflow to proceed without further platform-specific logic.

# Appendix C. Agent prompt specifications

## C.1 Prompt for Agent 0

The Coordinator Agent is initialized by a dedicated prompt that casts it as a “public opinion analysis operator” and instructs it to extract exactly three parameters,event keywords, start and end dates, and social media platform,from any free form user query.

At startup, this prompt (stored as a plain-text file) is loaded into the agent’s conversation history as its message, establishing its role and parsing rules. When a user submits a natural language query, that query is appended to the context and sent to the LLM, whose JSON response is then parsed by the agent into a structured instruction set. In this way, the Coordinator Agent seamlessly transforms a single user utterance into the precise parameters needed for downstream data collection and analysis.

The data collection utility inside Agent 0 is not driven by any prompt. Instead, once the Coordinator Agent has parsed the user’s query into a structured JSON object, it passes those parameters directly to its built-in crawler module. That module issues HTTP requests to the target platform, collects the matching posts based on those parameters, and writes the resulting records into a CSV file. No LLM invocations or prompt parsing occur during this step,everything is handled by standard program logic.

## C.2 Prompt for Agent 1

The Sentiment Analysis Agent is initialized with a dedicated prompt that frames its role as a classifier of public opinion posts.

At startup, the agent loads this plain-text prompt,directing it to assess each entry from the collected data CSV as positive, neutral, or negative,and inserts it as its message. When invoked, the agent combines this prompt with the batch of post texts and expects a JSON structured response containing both per-post sentiment labels and an overall sentiment summary.

By centralizing all classification rules and output formatting in its prompt, the Sentiment Analysis Agent can consistently and autonomously transform raw text into structured sentiment insights.

## C.3 Prompt for Agent 2

The Topic Extraction Agent is initialized with a dedicated prompt that defines its role as a topic identifier for public opinion posts.

On startup, the agent loads this plain-text prompt,directing it to analyze each entry from the collected data CSV, identify one or more salient topics, and return the results in a structured JSON format,and inserts it as its message. When invoked, the agent combines this prompt with the batch of post texts and parses the model's response into a list of topic labels for each post.

By encapsulating all topic extraction logic and output schema within its prompt, the Topic Extraction Agent reliably translates raw text into structured topic insights without any additional procedural code.

## C.4 Prompt for Agent 3

Agent 3 is responsible for synthesizing both sentiment and topic outputs into a final report, so it is driven by two distinct prompts,one for summarizing sentiment results and another for summarizing topic results,each tailored to its respective summarization role.

## C.4.1 Sentiment summarizer prompt

The first prompt instructs Agent 3 to transform raw sentiment counts into a concise narrative.

At startup, the agent loads this plain-text prompt,specifying that it should calculate percentages for positive, neutral, and negative categories and highlight key trends,and uses it as its message. When invoked, the agent combines the prompt with the sentiment CSV payload and parses the model's response into a polished summary paragraph.

## C.4.2 Topic summarizer prompt

The second prompt directs Agent 3 to generate a descriptive discussion of the extracted topics.

This prompt tells the model to take the list of topics and their frequencies, compute relative prominence, and articulate which topics are most significant and why. Upon invocation, the agent pairs the prompt with the topic-frequency CSV and captures the resulting narrative as a structured text block.

To clearly illustrate each prompt's function and its associated agent, Table 1 summarizes the role, prompt type, and output for each of the four agents, providing a concise overview of the pipeline’s modular structure.

**Table C.1 Summary of agents and prompts**

| Agent | Role | Prompt Type | Output |
|---|---|---|---|
| **Agent 0** | Instruction parsing & data collection | Dedicated prompt (plain-text file) | Structured JSON (“event_keywords”, “start_date”, “end_date”, “social_platform”); Collected_data.csv |
| **Agent 1** | Sentiment classification | Sentiment analysis prompt (plain-text file) | Sentiment_analysis_result.csv (per-post labels + summary) |
| **Agent 2** | Topic extraction | Topic extraction prompt (plain-text file) | Topic_extraction_result.csv (topic lists per post) |
| **Agent 3** | Report generation | Sentiment & topic summarizer prompts (plain-text file) | Final narrative report |